\documentclass[fleqn,twoside]{article}
\usepackage[headings]{espcrc2}

\readRCS
$Id: espcrc2.tex,v 1.2 2004/02/24 11:22:11 spepping Exp $
\usepackage{graphicx}
\usepackage[figuresright]{rotating}

\newcommand{\AmS}{{\protect\the\textfont2
  A\kern-.1667em\lower.5ex\hbox{M}\kern-.125emS}}

\title{Studying High $p_T$ Muons in Cosmic-Ray Air Showers}

\author{Spencer R. Klein\address[LBNL]{Nuclear Science Division,
Lawrence Berkeley National Laboratory, Berkeley, CA, 94720, USA}
        \thanks{Email: srklein@lbl.gov}
	}

\runtitle{High $p_T$ Muons}
\runauthor{S. Klein}

\begin{document}

\begin{abstract}

Most cosmic-ray air shower arrays have focused on detecting
electromagnetic shower particles and low energy muons.
A few groups (most notably MACRO + EASTOP and SPASE + AMANDA) have studied the high energy
muon component of showers.  However, these experiments had small solid
angles, and did not study muons far from the core. The IceTop + IceCube combination, with its 1 
km$^2$ muon detection area
can study muons far from the shower core.  IceCube can measure their energy loss ($dE/dx$), 
and hence their energy. With the
energy, and the known distribution of production heights, the transverse momentum
($p_T$) spectrum of high $p_T$ muons can be determined.  The production of these
muons is calculable in perturbative QCD, so the
measured muon spectra can be used to probe the composition of incident
cosmic-rays.

\end{abstract}

\maketitle

\section{Introduction}

Despite much effort, the composition of high-energy (above $10^{15}$ eV)
cosmic-rays is poorly known.  Satellite and balloon experiments have insufficient
exposure for high-statistics measurements.  Terrestrial experiments have used the
electromagnetic  and low energy muon components of air showers for
composition measurements.  The ratio of electromagnetic to muon energy may be sensitive to
the cosmic-ray composition, but it is also sensitive to the physics models used to simulate 
the air shower.  One critical component of these models is the forward production of
hadrons in high-energy interactions.  Most of these hadrons are
produced at low transverse momentum ($p_T$), where perturbative QCD (pQCD) is not applicable.
A variety of non-pertubative calculations and extrapolations are used to model
low $p_T$ particle production, and thereby make predictions about cosmic-ray composition.

Here, we present a complementary approach to study air showers, using high energy, high $p_T$ muons 
to study cosmic-ray composition.  The calculations use pQCD, and so should be well understood.
High $p_T$ muons are produced predominantly in the initial 
cosmic-ray interaction \cite{prompt2}, from semileptonic decays of heavy quarks, and from decays of 
high $p_T$ pions and kaons produced in jet fragmentation.  The rates
for muons with $p_T$ above a few GeV is calculable in pQCD; the spectrum depends on the parton 
composition of the incident ions.  For a fixed shower energy, the energy per nucleon
drops as the atomic number rises, substantially altering the parton density, and so changing
the muon spectrum.  This composition
may be inferred from the high $p_T$ muon spectrum.  The low$-x$ parton distributions in 
nitrogen also contribute in some kinematic areas; this may be an additional study topic.

\section{Experimental Technique}

The study of high $p_T$ muons requires a surface air shower array combined with a large 
underground muon detector.  The surface array measures the shower energy, core position and 
incident direction, and the underground detector
measures the energy and position of high-energy muons.  Previous experiments have
studied high energy muons in air showers, but, with relatively small underground detectors
\cite{MACRO}\cite{SPASE}.  These
experiments did not make use of the distance between the muon and the air shower core.  

IceCube and IceTop comprise a 1 km$^2$ surface air shower array
and a 1 km$^3$ muon detector \cite{ice}.  The muon detector is big enough to
observe muons far from the shower core. Together, the combination can determine
the key elements of the event.

IceTop
will measure the shower energy, 
core position, and arrival direction.
It will consist of 160 ice-filled tanks spread over  a 1 km$^2$ area\cite{bai}.  It has an energy
threshold of about 300 TeV.

IceCube will consist of up to 4800 optical modules in 1 km$^3$ of ice, at
depths from 1450 to 2450 meters.  The combined acceptance will be about 0.3 km$^2$ sr \cite{bai}. 
IceCube will measure the energy, position and direction of muons.  For vertical muons, 
the energy threshold is about 500 GeV
Muon energy, $E_\mu$ is measured by determining the muon specific energy loss ($dE/dx$). For $E_\mu > 1$
TeV, $dE/dx$ scales with $E_\mu$.

In a sufficiently high energy event, IceCube will observe multiple muons.  Most of these
muons will have low $p_T$, and so will cluster around the shower core.
In this high-density region, it may not be possible to reconstruct individual muons;
the bundle will be reconstructed as a single light source.  
However, far from the core, where the muon density is
lower, it should be possible to reconstruct individual tracks.
The separation required to resolve individual tracks is unknown,
but 100 meters (comparable to the spacing between optical module strings) 
seems like a safe value. It is more than 3 times the effective light scattering length 
of about 30 m \cite{icepaper}.

With the muon energy and distance from the core determined, the muon $p_T$ may be calculated:
\begin{equation}
p_T = {E_\mu d_c \over h}
\end{equation}
where $E_\mu$ is the muon energy, $d_c$ is its distance from the core, and $h$ is the
distance from IceCube to the site of the initial cosmic-ray interaction in the atmosphere.
The value of $h$ is not determined on an event-by-event basis, but its average
value ($\approx$ 30 km), zenith angle dependence and distribution are well known; the 
event-to-event variation and slight composition dependence can be considered
in determining the $p_T$ spectrum.

With the conservative $d_c> 100$ m requirement and a 1 TeV muon,
IceCube can study the spectrum for $p_T > 3$ GeV/c, covering most of the 
interesting high $p_T$ region.  As $E_\mu$
increases, so does the minimum accessible $p_T$; for $E_\mu=10$ TeV,  
$p_T > 30$ GeV/c, while for a 50 
TeV muon, $p_T > 150$ GeV/c.   Since higher energy muons produce more Cherenkov light, they are
easier to track, and it is likely that, for higher energy muons, a smaller $d_c$ cut could be used.

The systematic errors in these measurements remain to be studied.  Some components are:

1)  Uncertainty in the absolute cosmic-ray flux and energy scale.  The flux factors out, 
since we will only use
events observed by IceTop.  The energy scale introduces an uncertainty in the scale of
$x$ measurements.

2)  Error on the core position and extrapolation to depth.  For muons
far from the core, the fractional error is small.  In the first IceCube string,
the systematic offset between IceCube and IceTop is less than 1 degree \cite{performance}.

3)  Uncertainty in the muon energy and position, due to 
stochastic interactions, multiple scattering and other factors.
Multiple scattering contributes to $d_c$, but, far from the core, 
$d_c$ is dominated by the muon $p_T$.  

\section{Rates}

Thunman, Ingelman and Gondolo (TIG) calculated the prompt (charm only) and non-prompt muon
production using a PYTHIA Monte Carlo simulation, with the MRS-G parton distribution functions, 
leading-order pQCD cross-sections for $q\overline q\rightarrow Q\overline Q$ and
$gg\rightarrow Q\overline Q$, and standard charm quark hadronization and decay 
models \cite{prompt}.
Particles were propagated in the atmosphere using transport equations.  
Table \ref{t1} shows the expected muon rate in the
combined IceCube -IceTop acceptance for different energy thresholds.  Calculations,
with a fixed $E_\mu^{-3.7}$ energy spectrum find more non-prompt muons at low energies \cite{gaisser}.

\begin{table}[tb]
\caption{Muon rates for the TIG95 calculation, for 1 year
($3\times10^7$ s) with 0.3 km$^2$ sr acceptance.}
\label{t1}
\begin {tabular}{lrr}
\hline
Energy  &  Prompt & Non-Prompt \\
Threshold &  Rate & Rate \\
\hline
$10^{15}$ eV   & 1.5 & 0.5 \\
$10^{14}$ eV   & 1,000 & 18,500 \\
$10^{13}$ eV   & 56,000 & $10^{7}$ \\
$10^{12}$ eV   & 600,000 & $7\times10^8$ \\
\hline
\end{tabular}
\end{table}

A newer calculation uses next-to-leading-order pQCD calculations and the 
CTEQ3 parton distributions.  It finds prompt rates that are comparable to the
earlier calculation at 1 TeV,
but are higher at higher energies; at 100 TeV, the prompt cross section is about 8 times higher
\cite{prompt2}.  The difference is due largely to the different low$-x$ behavior of the
two parton distributions.

Bottom quark production was not included.  Calculations for the LHC
(at a comparable energy) find that, for muon $p_T >2$ GeV/c, bottom contributes
a larger muon signal than charm \cite{LHC}. 

For $E_\mu > 50$ TeV, the accompanying shower
should almost always trigger IceTop; at lower muon energies, 
some of the showers may be too small to trigger IceTop, so will not be seen.  

The prompt signal is significant
for $E_\mu < 10^{14}$ eV, although smaller than the non-prompt signal.  
A $p_T$ cut should eliminate all of the soft (non-perturbative) non-prompt signal, leaving muons from
heavy quarks and high $p_T$ $\pi$ and $K$. Based on calculations for
the LHC, a cut on $p_T>2-4$ GeV/c will also eliminate 90-98\% of the prompt muons\cite{LHC}. 
Although a drastic reduction, this still leaves an interesting sample.   The final cut on 
$p_T$ (or $d_c$) will depend on the detector 2-track (core + distant muon) 
separation capability.

\section{Muon spectrum Analysis}

Measurement of the muon $p_T$ spectra has some similarities with the lepton
spectra studies done at the Relativistic Heavy Ion Collider (RHIC).  The
$p_T$ spectrum of leptons produced in proton-proton, deuteron-gold and heavy-ion collisions
have been studied.

\begin{figure}[tb]
\center{\includegraphics[width=2.1 in,clip]{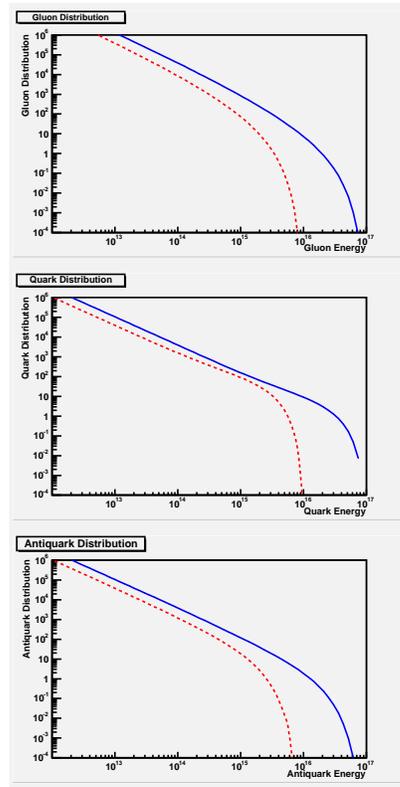}}
\vskip -0.2 in
\caption{Parton distribution for a $10^{17}$ eV cosmic ray for quarks (top), 
gluons (middle) and antiquarks (bottom). The solid lines are for
hydrogen ($A=1$), while the dashed lines are for $A=10$ nuclei. These
curves are based on the MRST99 parton distributions \cite{partons} 
at $Q^2 = 1000$ GeV$^2$.}
\label{fig:partons}
\end{figure}

The RHIC experiments fit their data using a multi-component 'cocktail.'  For electrons,
the cocktail consists of $\pi^0$ and $\eta$ Dalitz decays, $\gamma\rightarrow e^+e^-$
conversions, leptonic decays of vector mesons, plus semileptonic decays of heavy mesons and
baryons \cite{RHICe}. For muons, the cocktail consists of $\pi$, $K$ and heavy quark decays \cite{RHICmu}. 
For muons, the $\pi$ and $K$ decay fraction is reduced with vertex cuts,
sample, so this result is not directly relevant to IceCube.  However, the electron
analysis seems quite relevant.  The fraction of 
prompt electrons rises with the electron $p_T$; for $p_T >5$ GeV/c, prompt electrons are dominant. 
Additional confidence in these studies comes from the good agreement seen between high $p_T$ $\pi^0$
data and pQCD calculations\cite{pi0}.  

Simple arguments predict this dominance.  Light meson production is predominantly
soft; $d\sigma/dp_T$ falls exponentially with $\langle p_T\rangle \approx m$, $m$ being the meson
mass.
In contrast, heavy quark production is described by pQCD, which
gives a power law $p_T$ spectrum; at high enough $p_T$, this will dominate over any exponential.
pQCD processes also produce
high $p_T$ $\pi$ and K, but these mesons are only a small fraction of the total production.

A similar approach could apply to the $p_T$ spectrum of muons from air showers, with a cocktail
of perturbative and non-perturbative non-prompt muons, and prompt muons.  For $p_T$ above a 
few GeV/c, the non-perturbative component will be gone, and the remaining events could be fit to a 
power law spectrum.  Spectra could be measured for different muon energies and shower energies.
The muon energies are related to the muon rapidity in the center-of-mass frame. 

\section{Composition Determination}

Fig. 1 compares the parton energy spectrum (in the target frame) for a $10^{17}$ eV cosmic ray, for
hydrogen ($A=1$) and an $A=10$ nucleus.  The maximum parton energy scales as $1/A$; this determines
the maximum parton-parton center of mass energy. 

For a given collision, the $x$ values of the parton densities that are probed depend on the kinematics
of the produced partons.  So, different muon energies and $p_T$ are sensitive to different
$x$ regions.   

In any collision, the maximum possible muon energy is the maximum incident parton energy.  
The maximum $p_T$ is half the parton-parton center-of mass energy,
$W = \sqrt{2E x_P x_N m_p}$.
Here $E$ is cosmic-ray energy, $x_T$ and $x_N$ are the $x$ of projectile and target partons, 
and $m_p$ is the mass of the proton.  The corresponding spectra are determined by a calculation
that includes the kinematics of the parton production, fragmentation, and semileptonic decays. 

It may be worth giving a few examples of the range of $x$ values probed, using a
a grossly simplified model where the muon takes
half of the energy and $p_T$ of the parton produced in the collisions. 
For an incident $10^{18}$ eV proton producing a $10^{16}$ eV muon
with $p_T = 1$ GeV (this muon would be in the core, and only distinguishable due to
its huge $dE/dx$), $x_P$ = 0.01, $x_N=1.5\times10^{-6}$ and $Q^2\approx 30$ GeV$^2$.  With the same
incident particle and muon energy, a muon with $p_T = 100$ GeV would come from a collision
with $x_P=10^{-3}$, $x_N = 8\times10^{-2}$ and $Q^2=10^5$ GeV$^2$.  

\section{Conclusions}

Studies of high $p_T$ muon production in cosmic-ray air showers appears feasible with
IceCube and IceTop combined.  IceTop can measure the air shower energy, incident direction, and core position, while IceCube
will measure muon energy and distance from the shower core; these data can be used to determine the
muon $p_T$.  For $p_T$ above a few GeV, the muon $p_T$ spectrum can be interpreted in terms of perturbative
QCD plus and fragmentation functions.  The muon $p_T$ spectrum should be sensitive to the composition of
incident cosmic rays. 

It is a pleasure to acknowledge useful comments from Xinhua Bai and Tom Gaisser and
Teresa Montaruli.

\end{document}